\documentclass[aps,pra,twocolumn,amsmath,showpacs,showkeys,superscriptaddress,floatfix]{revtex4}

\usepackage{amsmath}
\usepackage{epsfig}
\usepackage{graphicx}
\usepackage{graphics}
\usepackage{url}
\usepackage[english]{babel}
\usepackage{dcolumn}
\usepackage{color}
\usepackage[squaren]{SIunits}
\begin{document}
\title{Phonon Effects on Population Inversion in Quantum Dots:
Resonant, Detuned and Frequency-swept Excitations}

\author{D.~E.~Reiter}
\affiliation{Institut f\"ur Festk\"orpertheorie, Universit\"at M\"unster,
Wilhelm-Klemm-Str.~10, 48149 M\"unster, Germany}

\author{S.~L\"uker}
\affiliation{Institut f\"ur Festk\"orpertheorie, Universit\"at M\"unster,
Wilhelm-Klemm-Str.~10, 48149 M\"unster, Germany}

\author{K.~Gawarecki}
\affiliation{Institute of Physics, Wroc{\l}aw University of Technology,
50-370 Wroc{\l}aw, Poland}

\author{A.~Grodecka-Grad}
\affiliation{QUANTOP, Danish National Research Foundation Center for Quantum
Optics, Niels Bohr Institute, University of Copenhagen, 2100 Copenhagen {\O},
Denmark}

\author{P.~Machnikowski}
\affiliation{Institute of Physics, Wroc{\l}aw University of Technology,
50-370 Wroc{\l}aw, Poland}

\author{V.~M.~Axt}
\affiliation{Theoretische Physik III, Universit\"at Bayreuth, 95440 Bayreuth,
Germany}

\author{T.~Kuhn}
\affiliation{Institut f\"ur Festk\"orpertheorie, Universit\"at M\"unster,
Wilhelm-Klemm-Str.~10, 48149 M\"unster, Germany}

\date{\today}

\begin{abstract}
The effect of acoustic phonons on different light-induced excitations of a semiconductor quantum dot is investigated. Resonant excitation of the quantum dot leads to Rabi oscillations, which are damped due to the phonon interaction. When the excitation frequency is detuned, an occupation can only occur due to phonon absorption or emission processes. For frequency-swept excitations a population inversion is achieved through adiabatic rapid passage, but the inversion is also damped by phonons. For all three scenarios the influence of the phonons depends non-monotonically on the pulse area.
\end{abstract}

\pacs{78.67.Hc; 78.47.D-; 42.50.Md}

\keywords{quantum dot; optical control; phonons}

\maketitle

\section{Introduction}

Phonons play an important role in the control of the quantum state of a semiconductor quantum dot (QD). When the QD is excited by a laser pulse, the system dynamics is affected by the electron-phonon interaction. In the most common case of an excitation resonant on the QD transition, the occupation of the QD exciton oscillates as function of the pulse area. This Rabi oscillation is damped by the interaction of the exciton system with phonons, which has been analyzed theoretically \cite{forstner2003pho,machnikowski2004res,krugel2005the,vagov2007non,glassl2011lon,mccutcheon2010qua,mccutcheon2011gen} and experimentally observed \cite{zrenner2002coh,ramsay2010dam,ramsay2010pho}. A particularly remarkable feature is that the damping is non-monotonic, for sufficiently large pulse areas a reappearance of the Rabi oscillations has been predicted \cite{vagov2007non}. The achieved population inversion by resonant optical excitation is sensitive to the pulse area. For small deviations from the resonance condition slightly modified oscillations appear \cite{ramsay2011eff}. For large detunings no population inversion takes place without phonons. In this case only through emission or absorption of phonons an occupation becomes possible. To achieve a population inversion which is stable against small variations of the excitation conditions also the adiabatic rapid passage (ARP) can be used, which has recently been demonstrated experimentally \cite{simon2011rob,wu2011pop}. When excited by a chirped laser pulse, the system follows the eigenstates of the coupled system-light Hamiltonian, i.e., the dressed states, adiabatically. Due to the chirp the dressed states change their character during the pulse, which results in a population inversion. Because phonons can induce transitions between the dressed states, a damping of the ARP effect is seen \cite{luker2012inf}.

In this paper, we study theoretically the phonon effects of an optically excited QD and compare the three excitation scenarios mentioned above. We analyze the phonon influence as a function of the pulse area and study the occupation at high pulse areas, where a reappearance is expected. Furthermore we show that for low temperatures phonon emission and absorption processes are not balanced and asymmetries with respect to either the detuning or the chirp appear. For high temperatures the behavior becomes symmetric.

\section{Theory}
For the calculations, we model the QD as a two-level system, which consists of the ground state $|0\rangle$ and the exciton state $|X\rangle$ split by the transition energy $\hbar \Omega_X$. The excitation of the QD is modeled by a circularly polarized light field coupled in the usual rotating wave and dipole approximations with instantaneous Rabi frequency $\Omega^{(+)}(t)=2ME^{(+)}(t)/\hbar$, $\Omega^{(-)}=\Omega^{(+)*}$, with the dipole matrix element $M$ and the positive (negative) frequency component of the light field $E^{(+)}$ ($E^{(-)}$). The phonon coupling is diagonal in the carrier states with the creation and annihilation operators $b_{\mathbf{q}}^\dag$ and $b_{\mathbf{q}}^{}$. Because the coupling to longitudinal acoustic (LA) phonons has been shown to be most important for dephasing in typical InGaAs QD structures \cite{vagov2004non,krummheuer2005pur}, we restrict ourselves to LA phonon with the dispersion relation $\omega_{\mathbf{q}} = cq$, $c$ being the sound velocity. The phonons couple to the QD exciton via the coupling matrix element $g_{\mathbf{q}}$. We assume the QD to be spherical with a size of $4$~nm and take GaAs material parameters \cite{krummheuer2002the,krugel2006bac}. The Hamiltonian of the system is
\begin{eqnarray} \nonumber
H &=& \hbar\Omega_X |X\rangle\langle X| - \frac{\hbar\Omega^{(+)}(t)}{2}
|0\rangle\langle X| - \frac{\hbar\Omega^{(-)}(t)}{2} |X\rangle\langle 0| \\
&& + \hbar \sum_{\mathbf{q}}\omega_{\mathbf{q}}b_{\mathbf{q}}^\dag
b_{\mathbf{q}} + \hbar
\sum_{\mathbf{q}}\left(g_{\mathbf{q}}b_{\mathbf{q}} +
g_{\mathbf{q}}^* b_{\mathbf{q}}^\dag\right)  |X\rangle\langle X|.
\end{eqnarray}
Initially the system is in the ground state $|0\rangle$ and the phonons are in a thermal state with temperature $T$. The equations of motion are set up by the Heisenberg equation, which leads to the well known infinite hierarchy of equations due to the many-body interaction. This hierarchy is truncated by a correlation expansion in fourth order, which has been shown to perfectly agree with a numerically exact path integral method in a wide range of parameters \cite{glassl2011lon}.

In this paper we will consider different light excitation scenarios. On the one hand we look at excitation with a laser pulse with a constant frequency $\omega_{L}$. With a Gaussian envelope of the duration $\tau$ and an amplitude $E_0$ the pulse reads $ E^{(+)}(t) = E_0 e^{-\frac{t^2}{2\tau^2}} e^{-i\omega_{L} t}$. The laser pulse frequency $\omega_{L}$ is either set resonant to the polaron shifted transition energy or it is detuned by a given frequency shift $\Delta/\hbar$. On the other hand we consider an excitation with a frequency swept laser pulse $E^{(+)}(t) = E_0 e^{-\frac{t^2}{2\tau^2}} e^{-i(\omega_{L}+\frac{1}{2} a t) t}$, where the frequency changes linearly with time by the chirp $a$. The laser pulse duration is kept fixed to $\tau=4$~ps for all pulses.

\section{Results}
\subsection{Constant laser frequency}

\begin{figure}[htb]
        \includegraphics[width=0.70\columnwidth]{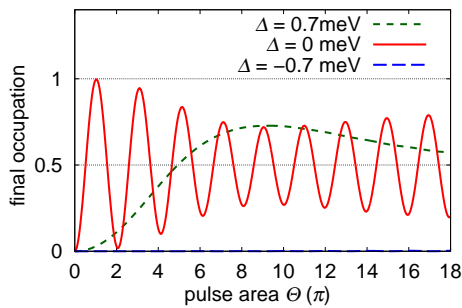}
    \caption{(Color online) Occupation of the exciton state as function of the pulse area $\Theta$ for resonant excitation $\Delta=0$ and detuned excitation $\Delta = \pm0.7$~meV at temperature $T=1$~K.}
        \label{fig:rabi}
\end{figure}

When a QD is excited by a laser pulse, after the pulse the exciton occupation does not change due to phonon processes anymore. Thus, a good quantity to look at is the final occupation after the pulse as function of the laser pulse area $\Theta=\frac{2}{\hbar}\int |M E(t)|dt$, which is proportional to the square root of the laser intensity. This is shown in Figure~\ref{fig:rabi} for resonant excitation and detuned excitation with $\Delta=\pm 0.7$~meV. The temperature is $T=1$~K. For resonant excitation, $\Delta =0$, clearly Rabi oscillations are seen, which are damped by the electron-phonon interaction. For higher pulse areas the amplitude of the Rabi oscillations is increased again, which is referred to as reappearance \cite{vagov2007non}. This behavior can be understood qualitatively by comparing the Rabi frequency to the frequency of the phonons. For low pulse areas the laser induced Rabi oscillation is much slower than the phonon dynamics and the phonons follow adiabatically. For high pulse areas the Rabi oscillation is much faster than the phonons, which cannot follow anymore. Inbetween the optically induced oscillation and the phonon dynamics are in resonance, which is where the maximal damping is found. The details of the reappearance depend on the pulse shape \cite{glassl2011inf}. For our parameters, the maximal damping is achieved around $\Theta=9\pi$ as can be seen in Fig.~\ref{fig:rabi}. The quantitative value of the maximal damped pulse area depends on the QD size and the pulse length.

Figure~\ref{fig:rabi} also shows the occupation for a detuning of $\Delta=\pm0.7$~meV. For a negative detuning $\Delta=-0.7$~meV the occupation is zero for all pulse areas. For positive detuning we see that the occupation rises with increasing pulse area to a maximum of $0.75$ around $\Theta=9\pi$ and then decreases slightly. Here phonons are emitted during the excitation process, which leads to an occupation of the exciton state.

\begin{figure}[htb]
        \includegraphics[width=1.0\columnwidth]{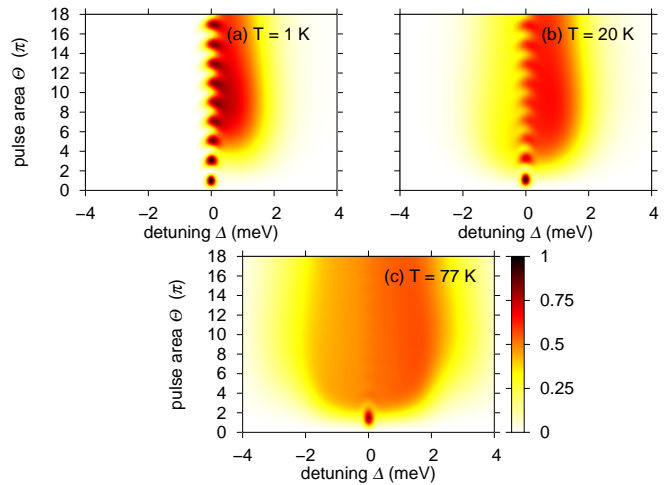}
    \caption{(Color online) Contour plot of the final occupation for detuned excitation as function of detuning $\Delta$ and pulse area $\Theta$ for temperatures (a) $T=1$~K, (b) $T=20$~K, and (c) $T=77$~K. }
        \label{fig:detune}
\end{figure}

In Fig.~\ref{fig:detune} a contour plot of the final occupation as a function of detuning and pulse area is shown. At $T=1$~K (Fig.~\ref{fig:detune}(a)) for resonant excitation, i.e. $\Delta=0$, the Rabi oscillations are seen. For small detuning $|\Delta|<0.4$~meV off-resonant Rabi oscillations are observed. For large detunings the occupation after excitation should be zero if no phonons are included. Indeed, in Fig.~\ref{fig:detune}(a) for negative detuning the occupation is zero. For positive detuning we find that a non-zero occupation up to $0.75$ for detunings up to $\Delta=1.5$~meV. As can be seen in Fig.~\ref{fig:detune}(a), the phonon-assisted exciton generation is most effective around $\Theta=9\pi$, mostly independent of the value of the detuning.

For higher temperatures (Fig.~\ref{fig:detune}(b),(c)) phonon absorption processes become also effective. As shown in Fig.~\ref{fig:detune}(b) at $T=20$~K, for negative detunings an occupation of the exciton state can be achieved by absorbing a phonon. The resonant Rabi oscillations are damped more efficiently as now more phonons are available for interaction. For positive detunings, absorption processes also take place and reduce the occupation as compared to the $1$~K temperature case. At a still higher temperature of $T=77$~K, shown in Fig.~\ref{fig:detune}(c), we find that absorption and emission processes are nearly balanced and an occupation of about $0.5$ is reached for resonant excitation and detunings up to $|\Delta|<3$~meV.

\subsection{Chirped laser pulse}

\begin{figure}[htb]
        \includegraphics[width=1.\columnwidth]{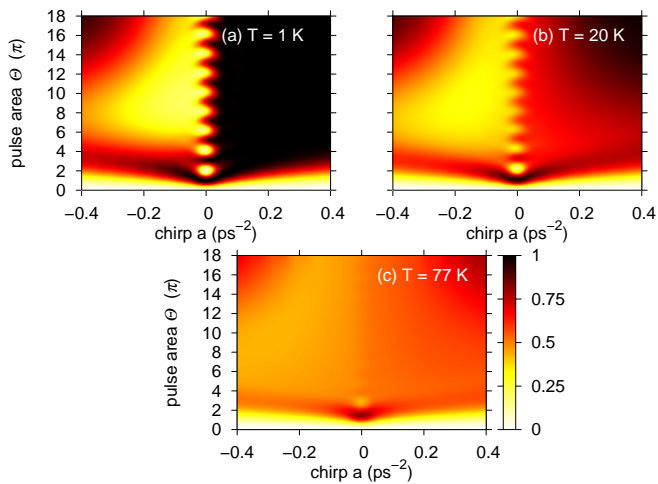}
    \caption{(Color online) Contour plot of the final occupation for chirped excitation for temperatures (a) $T=1$~K, (b) $T=20$~K, and (c) $T=77$~K. }
        \label{fig:chirp}
\end{figure}

If a chirped pulse excites the system, the ARP provides an efficient mechanism for a population inversion. Without phonons a population inversion is achieved as soon as the threshold for the adiabatic regime is reached. When the system interacts with phonons, a damping of the population inversion is seen \cite{luker2012inf}. Figure~\ref{fig:chirp} shows a contour plot of the occupation as function of chirp and pulse area for three different temperatures $T=1,20$, and $77$~K. At low temperatures ($T=1$~K in Fig.~\ref{fig:chirp}(a)) only phonon emission processes can occur. Accordingly, only for negative chirps a damping of the ARP is seen, while for positive chirps the occupation reaches one. Also in this case the phonon coupling depends non-monotonically on the pulse area. The highest damping is found again around $\Theta=9\pi$, where the occupation is close to zero for negative chirps. For high pulse areas the occupation rises again, in particular for high chirps. Here, the time when the optical driving and the phonon dynamics are in resonance is even shorter due to the change in frequency.

For higher temperatures phonon absorption becomes more likely and even balanced with phonon emission as seen in Fig.~\ref{fig:chirp}(b) and (c). At $T=20$~K, for positive chirps a full inversion is not reached anymore. For $T=77$~K, the occupation is about $0.5$ for a wide range of pulse areas and chirps.

\section{Conclusions}

In summary, we have studied the impact of phonons on the final occupation of a QD exciton excited by a resonant, detuned or chirped laser pulse. For resonant and chirped excitation, phonons hinder the population inversion, while for detuned excitation an occupation becomes possible by phonon interaction. For all scenarios, the phonon influence depends non-monotonically on the pulse area, leading to a maximal phonon effect around a pulse area of $\Theta=9\pi$. Because phonon absorption is unlikely at low temperatures, asymmetries with respect to the detuning and the chirp, respectively, occur, which diminish at higher temperatures.

\textbf{Acknowlegdements:}
This work was supported by a Research Group Linkage Project of the
Alexander von Humboldt Foundation and by the TEAM programme of the Foundation
for Polish Science co-financed from the European Regional Development Fund.


\begin{thebibliography}{19}
\expandafter\ifx\csname natexlab\endcsname\relax\def\natexlab#1{#1}\fi
\expandafter\ifx\csname bibnamefont\endcsname\relax
  \def\bibnamefont#1{#1}\fi
\expandafter\ifx\csname bibfnamefont\endcsname\relax
  \def\bibfnamefont#1{#1}\fi
\expandafter\ifx\csname citenamefont\endcsname\relax
  \def\citenamefont#1{#1}\fi
\expandafter\ifx\csname url\endcsname\relax
  \def\url#1{\texttt{#1}}\fi
\expandafter\ifx\csname urlprefix\endcsname\relax\def\urlprefix{URL }\fi
\providecommand{\bibinfo}[2]{#2}
\providecommand{\eprint}[2][]{\url{#2}}

\bibitem[{\citenamefont{Gl{\"a}ssl
  et~al.}(2011{\natexlab{a}})\citenamefont{Gl{\"a}ssl, Vagov, L{\"u}ker,
  Reiter, Croitoru, Machnikowski, Axt, and Kuhn}}]{glassl2011lon}
\bibinfo{author}{\bibfnamefont{M.}~\bibnamefont{Gl{\"a}ssl}},
  \bibinfo{author}{\bibfnamefont{A.}~\bibnamefont{Vagov}},
  \bibinfo{author}{\bibfnamefont{S.}~\bibnamefont{L{\"u}ker}},
  \bibinfo{author}{\bibfnamefont{D.~E.} \bibnamefont{Reiter}},
  \bibinfo{author}{\bibfnamefont{M.~D.} \bibnamefont{Croitoru}},
  \bibinfo{author}{\bibfnamefont{P.}~\bibnamefont{Machnikowski}},
  \bibinfo{author}{\bibfnamefont{V.~M.} \bibnamefont{Axt}}, \bibnamefont{and}
  \bibinfo{author}{\bibfnamefont{T.}~\bibnamefont{Kuhn}},
  \bibinfo{journal}{Phys.\ Rev.\ {\rm B}} \textbf{\bibinfo{volume}{84}},
  \bibinfo{pages}{195311} (\bibinfo{year}{2011}{\natexlab{a}}).

\bibitem[{\citenamefont{McCutcheon and Nazir}(2010)}]{mccutcheon2010qua}
\bibinfo{author}{\bibfnamefont{D.~P.~S.} \bibnamefont{McCutcheon}}
  \bibnamefont{and} \bibinfo{author}{\bibfnamefont{A.}~\bibnamefont{Nazir}},
  \bibinfo{journal}{New J.\ Phys} \textbf{\bibinfo{volume}{12}},
  \bibinfo{pages}{113042} (\bibinfo{year}{2010}).

\bibitem[{\citenamefont{F{\"o}rstner et~al.}(2003)\citenamefont{F{\"o}rstner,
  Weber, Danckwerts, and Knorr}}]{forstner2003pho}
\bibinfo{author}{\bibfnamefont{J.}~\bibnamefont{F{\"o}rstner}},
  \bibinfo{author}{\bibfnamefont{C.}~\bibnamefont{Weber}},
  \bibinfo{author}{\bibfnamefont{J.}~\bibnamefont{Danckwerts}},
  \bibnamefont{and} \bibinfo{author}{\bibfnamefont{A.}~\bibnamefont{Knorr}},
  \bibinfo{journal}{Phys.\ Rev.\ Lett.} \textbf{\bibinfo{volume}{91}},
  \bibinfo{pages}{127401} (\bibinfo{year}{2003}).

\bibitem[{\citenamefont{Kr{\"u}gel et~al.}(2005)\citenamefont{Kr{\"u}gel, Axt,
  Kuhn, Machnikowski, and Vagov}}]{krugel2005the}
\bibinfo{author}{\bibfnamefont{A.}~\bibnamefont{Kr{\"u}gel}},
  \bibinfo{author}{\bibfnamefont{V.~M.} \bibnamefont{Axt}},
  \bibinfo{author}{\bibfnamefont{T.}~\bibnamefont{Kuhn}},
  \bibinfo{author}{\bibfnamefont{P.}~\bibnamefont{Machnikowski}},
  \bibnamefont{and} \bibinfo{author}{\bibfnamefont{A.}~\bibnamefont{Vagov}},
  \bibinfo{journal}{Appl.\ Phys.\ B} \textbf{\bibinfo{volume}{81}},
  \bibinfo{pages}{897} (\bibinfo{year}{2005}).

\bibitem[{\citenamefont{Vagov et~al.}(2007)\citenamefont{Vagov, Croitoru, Axt,
  Kuhn, and Peeters}}]{vagov2007non}
\bibinfo{author}{\bibfnamefont{A.}~\bibnamefont{Vagov}},
  \bibinfo{author}{\bibfnamefont{M.~D.} \bibnamefont{Croitoru}},
  \bibinfo{author}{\bibfnamefont{V.~M.} \bibnamefont{Axt}},
  \bibinfo{author}{\bibfnamefont{T.}~\bibnamefont{Kuhn}}, \bibnamefont{and}
  \bibinfo{author}{\bibfnamefont{F.~M.} \bibnamefont{Peeters}},
  \bibinfo{journal}{Phys.\ Rev.\ Lett.} \textbf{\bibinfo{volume}{98}},
  \bibinfo{pages}{227403} (\bibinfo{year}{2007}).

\bibitem[{\citenamefont{McCutcheon et~al.}(2011)\citenamefont{McCutcheon,
  Dattani, Gauger, Lovett, and Nazir}}]{mccutcheon2011gen}
\bibinfo{author}{\bibfnamefont{D.~P.~S.} \bibnamefont{McCutcheon}},
  \bibinfo{author}{\bibfnamefont{N.~S.} \bibnamefont{Dattani}},
  \bibinfo{author}{\bibfnamefont{E.~M.} \bibnamefont{Gauger}},
  \bibinfo{author}{\bibfnamefont{B.~W.} \bibnamefont{Lovett}},
  \bibnamefont{and} \bibinfo{author}{\bibfnamefont{A.}~\bibnamefont{Nazir}},
  \bibinfo{journal}{Phys.\ Rev.\ {\rm B}} \textbf{\bibinfo{volume}{84}},
  \bibinfo{pages}{081305} (\bibinfo{year}{2011}).

\bibitem[{\citenamefont{Machnikowski and Jacak}(2004)}]{machnikowski2004res}
\bibinfo{author}{\bibfnamefont{P.}~\bibnamefont{Machnikowski}}
  \bibnamefont{and} \bibinfo{author}{\bibfnamefont{L.}~\bibnamefont{Jacak}},
  \bibinfo{journal}{Phys.\ Rev.\ {\rm B}} \textbf{\bibinfo{volume}{69}},
  \bibinfo{pages}{193302} (\bibinfo{year}{2004}).

\bibitem[{\citenamefont{Ramsay et~al.}(2010{\natexlab{a}})\citenamefont{Ramsay,
  Gopal, Gauger, Nazir, Lovett, Fox, and Skolnick}}]{ramsay2010dam}
\bibinfo{author}{\bibfnamefont{A.~J.} \bibnamefont{Ramsay}},
  \bibinfo{author}{\bibfnamefont{A.~V.} \bibnamefont{Gopal}},
  \bibinfo{author}{\bibfnamefont{E.~M.} \bibnamefont{Gauger}},
  \bibinfo{author}{\bibfnamefont{A.}~\bibnamefont{Nazir}},
  \bibinfo{author}{\bibfnamefont{B.~W.} \bibnamefont{Lovett}},
  \bibinfo{author}{\bibfnamefont{A.~M.} \bibnamefont{Fox}}, \bibnamefont{and}
  \bibinfo{author}{\bibfnamefont{M.~S.} \bibnamefont{Skolnick}},
  \bibinfo{journal}{Phys.\ Rev.\ Lett.} \textbf{\bibinfo{volume}{104}},
  \bibinfo{pages}{17402} (\bibinfo{year}{2010}{\natexlab{a}}).

\bibitem[{\citenamefont{Ramsay et~al.}(2010{\natexlab{b}})\citenamefont{Ramsay,
  Godden, Boyle, Gauger, Nazir, Lovett, Fox, and Skolnick}}]{ramsay2010pho}
\bibinfo{author}{\bibfnamefont{A.~J.} \bibnamefont{Ramsay}},
  \bibinfo{author}{\bibfnamefont{T.~M.} \bibnamefont{Godden}},
  \bibinfo{author}{\bibfnamefont{S.~J.} \bibnamefont{Boyle}},
  \bibinfo{author}{\bibfnamefont{E.~M.} \bibnamefont{Gauger}},
  \bibinfo{author}{\bibfnamefont{A.}~\bibnamefont{Nazir}},
  \bibinfo{author}{\bibfnamefont{B.~W.} \bibnamefont{Lovett}},
  \bibinfo{author}{\bibfnamefont{A.~M.} \bibnamefont{Fox}}, \bibnamefont{and}
  \bibinfo{author}{\bibfnamefont{M.~S.} \bibnamefont{Skolnick}},
  \bibinfo{journal}{Phys.\ Rev.\ Lett.} \textbf{\bibinfo{volume}{105}},
  \bibinfo{pages}{177402} (\bibinfo{year}{2010}{\natexlab{b}}).

\bibitem[{\citenamefont{Zrenner et~al.}(2002)\citenamefont{Zrenner, Beham,
  Stufler, Findeis, Bichler, and Abstreiter}}]{zrenner2002coh}
\bibinfo{author}{\bibfnamefont{A.}~\bibnamefont{Zrenner}},
  \bibinfo{author}{\bibfnamefont{E.}~\bibnamefont{Beham}},
  \bibinfo{author}{\bibfnamefont{S.}~\bibnamefont{Stufler}},
  \bibinfo{author}{\bibfnamefont{F.}~\bibnamefont{Findeis}},
  \bibinfo{author}{\bibfnamefont{M.}~\bibnamefont{Bichler}}, \bibnamefont{and}
  \bibinfo{author}{\bibfnamefont{G.}~\bibnamefont{Abstreiter}},
  \bibinfo{journal}{Nature} \textbf{\bibinfo{volume}{418}},
  \bibinfo{pages}{612} (\bibinfo{year}{2002}).

\bibitem[{\citenamefont{Ramsay et~al.}(2011)\citenamefont{Ramsay, Godden,
  Boyle, Gauger, Nazir, Lovett, Gopal, Fox, and Skolnick}}]{ramsay2011eff}
\bibinfo{author}{\bibfnamefont{A.~J.} \bibnamefont{Ramsay}},
  \bibinfo{author}{\bibfnamefont{T.~M.} \bibnamefont{Godden}},
  \bibinfo{author}{\bibfnamefont{S.~J.} \bibnamefont{Boyle}},
  \bibinfo{author}{\bibfnamefont{E.~M.} \bibnamefont{Gauger}},
  \bibinfo{author}{\bibfnamefont{A.}~\bibnamefont{Nazir}},
  \bibinfo{author}{\bibfnamefont{B.~W.} \bibnamefont{Lovett}},
  \bibinfo{author}{\bibfnamefont{A.}~\bibnamefont{Gopal}},
  \bibinfo{author}{\bibfnamefont{A.~M.} \bibnamefont{Fox}}, \bibnamefont{and}
  \bibinfo{author}{\bibfnamefont{M.~S.} \bibnamefont{Skolnick}},
  \bibinfo{journal}{J.\ Appl.\ Phys.} \textbf{\bibinfo{volume}{109}},
  \bibinfo{pages}{102415} (\bibinfo{year}{2011}).

\bibitem[{\citenamefont{Simon et~al.}(2011)\citenamefont{Simon, Belhadj,
  Chatel, Amand, Renucci, Lemaitre, Krebs, Dalgarno, Warburton, Marie
  et~al.}}]{simon2011rob}
\bibinfo{author}{\bibfnamefont{C.~M.} \bibnamefont{Simon}},
  \bibinfo{author}{\bibfnamefont{T.}~\bibnamefont{Belhadj}},
  \bibinfo{author}{\bibfnamefont{B.}~\bibnamefont{Chatel}},
  \bibinfo{author}{\bibfnamefont{T.}~\bibnamefont{Amand}},
  \bibinfo{author}{\bibfnamefont{P.}~\bibnamefont{Renucci}},
  \bibinfo{author}{\bibfnamefont{A.}~\bibnamefont{Lemaitre}},
  \bibinfo{author}{\bibfnamefont{O.}~\bibnamefont{Krebs}},
  \bibinfo{author}{\bibfnamefont{P.~A.} \bibnamefont{Dalgarno}},
  \bibinfo{author}{\bibfnamefont{R.~J.} \bibnamefont{Warburton}},
  \bibinfo{author}{\bibfnamefont{X.}~\bibnamefont{Marie}},
  \bibnamefont{et~al.}, \bibinfo{journal}{Phys.\ Rev.\ Lett.}
  \textbf{\bibinfo{volume}{106}}, \bibinfo{pages}{166801}
  (\bibinfo{year}{2011}).

\bibitem[{\citenamefont{Wu et~al.}(2011)\citenamefont{Wu, Piper, Ediger,
  Brereton, Schmidgall, Eastham, Hugues, Hopkinson, and Phillips}}]{wu2011pop}
\bibinfo{author}{\bibfnamefont{Y.}~\bibnamefont{Wu}},
  \bibinfo{author}{\bibfnamefont{I.~M.} \bibnamefont{Piper}},
  \bibinfo{author}{\bibfnamefont{M.}~\bibnamefont{Ediger}},
  \bibinfo{author}{\bibfnamefont{P.}~\bibnamefont{Brereton}},
  \bibinfo{author}{\bibfnamefont{E.~R.} \bibnamefont{Schmidgall}},
  \bibinfo{author}{\bibfnamefont{P.~R.} \bibnamefont{Eastham}},
  \bibinfo{author}{\bibfnamefont{M.}~\bibnamefont{Hugues}},
  \bibinfo{author}{\bibfnamefont{M.}~\bibnamefont{Hopkinson}},
  \bibnamefont{and} \bibinfo{author}{\bibfnamefont{R.~T.}
  \bibnamefont{Phillips}}, \bibinfo{journal}{Phys.\ Rev.\ Lett.}
  \textbf{\bibinfo{volume}{106}}, \bibinfo{pages}{067401}
  (\bibinfo{year}{2011}).

\bibitem[{\citenamefont{L{\"u}ker et~al.}(2012)\citenamefont{L{\"u}ker,
  Gawarecki, Reiter, Grodecka-Grad, Axt, Machnikowski, and
  Kuhn}}]{luker2012inf}
\bibinfo{author}{\bibfnamefont{S.}~\bibnamefont{L{\"u}ker}},
  \bibinfo{author}{\bibfnamefont{K.}~\bibnamefont{Gawarecki}},
  \bibinfo{author}{\bibfnamefont{D.~E.} \bibnamefont{Reiter}},
  \bibinfo{author}{\bibfnamefont{A.}~\bibnamefont{Grodecka-Grad}},
  \bibinfo{author}{\bibfnamefont{V.~M.} \bibnamefont{Axt}},
  \bibinfo{author}{\bibfnamefont{P.}~\bibnamefont{Machnikowski}},
  \bibnamefont{and} \bibinfo{author}{\bibfnamefont{T.}~\bibnamefont{Kuhn}},
  \bibinfo{journal}{Phys.\ Rev.\ {\rm B}} \textbf{\bibinfo{volume}{85}},
  \bibinfo{pages}{121302} (\bibinfo{year}{2012}).

\bibitem[{\citenamefont{Vagov et~al.}(2004)\citenamefont{Vagov, Axt, Kuhn,
  Langbein, Borri, and Woggon}}]{vagov2004non}
\bibinfo{author}{\bibfnamefont{A.}~\bibnamefont{Vagov}},
  \bibinfo{author}{\bibfnamefont{V.~M.} \bibnamefont{Axt}},
  \bibinfo{author}{\bibfnamefont{T.}~\bibnamefont{Kuhn}},
  \bibinfo{author}{\bibfnamefont{W.}~\bibnamefont{Langbein}},
  \bibinfo{author}{\bibfnamefont{P.}~\bibnamefont{Borri}}, \bibnamefont{and}
  \bibinfo{author}{\bibfnamefont{U.}~\bibnamefont{Woggon}},
  \bibinfo{journal}{Phys.\ Rev.\ {\rm B}} \textbf{\bibinfo{volume}{70}},
  \bibinfo{pages}{201305} (\bibinfo{year}{2004}).

\bibitem[{\citenamefont{Krummheuer et~al.}(2005)\citenamefont{Krummheuer, Axt,
  Kuhn, D'Amico, and Rossi}}]{krummheuer2005pur}
\bibinfo{author}{\bibfnamefont{B.}~\bibnamefont{Krummheuer}},
  \bibinfo{author}{\bibfnamefont{V.~M.} \bibnamefont{Axt}},
  \bibinfo{author}{\bibfnamefont{T.}~\bibnamefont{Kuhn}},
  \bibinfo{author}{\bibfnamefont{I.}~\bibnamefont{D'Amico}}, \bibnamefont{and}
  \bibinfo{author}{\bibfnamefont{F.}~\bibnamefont{Rossi}},
  \bibinfo{journal}{Phys.\ Rev.\ {\rm B}} \textbf{\bibinfo{volume}{71}},
  \bibinfo{pages}{235329} (\bibinfo{year}{2005}).

\bibitem[{\citenamefont{Krummheuer et~al.}(2002)\citenamefont{Krummheuer, Axt,
  and Kuhn}}]{krummheuer2002the}
\bibinfo{author}{\bibfnamefont{B.}~\bibnamefont{Krummheuer}},
  \bibinfo{author}{\bibfnamefont{V.~M.} \bibnamefont{Axt}}, \bibnamefont{and}
  \bibinfo{author}{\bibfnamefont{T.}~\bibnamefont{Kuhn}},
  \bibinfo{journal}{Phys.\ Rev.\ {\rm B}} \textbf{\bibinfo{volume}{65}},
  \bibinfo{pages}{195313} (\bibinfo{year}{2002}).

\bibitem[{\citenamefont{Kr{\"u}gel et~al.}(2006)\citenamefont{Kr{\"u}gel, Axt,
  and Kuhn}}]{krugel2006bac}
\bibinfo{author}{\bibfnamefont{A.}~\bibnamefont{Kr{\"u}gel}},
  \bibinfo{author}{\bibfnamefont{V.~M.} \bibnamefont{Axt}}, \bibnamefont{and}
  \bibinfo{author}{\bibfnamefont{T.}~\bibnamefont{Kuhn}},
  \bibinfo{journal}{Phys.\ Rev.\ {\rm B}} \textbf{\bibinfo{volume}{73}},
  \bibinfo{pages}{035302} (\bibinfo{year}{2006}).

\bibitem[{\citenamefont{Gl{\"a}ssl
  et~al.}(2011{\natexlab{b}})\citenamefont{Gl{\"a}ssl, Croitoru, Vagov, Axt,
  and Kuhn}}]{glassl2011inf}
\bibinfo{author}{\bibfnamefont{M.}~\bibnamefont{Gl{\"a}ssl}},
  \bibinfo{author}{\bibfnamefont{M.~D.} \bibnamefont{Croitoru}},
  \bibinfo{author}{\bibfnamefont{A.}~\bibnamefont{Vagov}},
  \bibinfo{author}{\bibfnamefont{V.~M.} \bibnamefont{Axt}}, \bibnamefont{and}
  \bibinfo{author}{\bibfnamefont{T.}~\bibnamefont{Kuhn}},
  \bibinfo{journal}{Phys.\ Rev.\ {\rm B}} \textbf{\bibinfo{volume}{84}},
  \bibinfo{pages}{125304} (\bibinfo{year}{2011}{\natexlab{b}}).

\end{thebibliography}

\end{document}